\documentclass[aps,prl,10pt,twocolumn,showpacs,superscriptaddress]{revtex4-1}
\usepackage{graphicx}
\usepackage{multirow}
\usepackage{color}
\usepackage{dcolumn}

\newcommand{\sgn}{\mathop{\mathrm{sgn}}}
\newcommand{\appsuppl}{Appendix}
\begin{document}

\title{Electron removal energies in noble gas atoms up to 100 keV: \textit{ab initio} $GW$ vs XPS}

\author{Iskander Mukatayev}
\affiliation{Universit\'e Grenoble Alpes, CEA, Leti, F-38000, Grenoble, France}
\author{Beno\^it Skl\'enard}
\affiliation{Universit\'e Grenoble Alpes, CEA, Leti, F-38000, Grenoble, France}
\affiliation{European Theoretical Spectroscopy Facility (ETSF)}
\author{Valerio Olevano}
\email{valerio.olevano@neel.cnrs.fr}
\affiliation{Universit\'e Grenoble Alpes, F-38000 Grenoble, France}
\affiliation{CNRS, Institut N\'eel, F-38042 Grenoble, France}
\affiliation{European Theoretical Spectroscopy Facility (ETSF)}
\author{Jing Li}
\email{jing.li@cea.fr}
\affiliation{Universit\'e Grenoble Alpes, CEA, Leti, F-38000, Grenoble, France}
\affiliation{European Theoretical Spectroscopy Facility (ETSF)}

\date{\today}

\begin{abstract}
X-ray photoelectron spectroscopy (XPS) measures electron removal energies, providing direct access to core and valence electron binding energies, hence probing the electronic structure.
In this work, we benchmark for the first time the \textit{ab initio} many-body $GW$ approximation on the complete electron binding energies of noble gas atoms (He-Rn), which spans 100~keV.
Our results demonstrate that $GW$ achieves an accuracy within 1.2\% in XPS binding energies, by systematically restoring the underestimation from density-functional theory (DFT, error of 14\%) or the overestimation from Hartree-Fock (HF, error of 4.7\%).
Such results also imply the correlations of $d$ electrons are very well described by $GW$.
\end{abstract}


\maketitle

\paragraph*{Introduction. ---}
The electronic structure of atoms, molecules and solids \cite{MartinReiningCeperley} is characterized by neutral excitations, as measured in optical spectroscopy, and charged electron removal/addition excitations, as measured in direct/inverse photoemission spectroscopy.
In X-ray photoelectron spectroscopy (XPS) \cite{Siegbahn,Carlson} an X-ray photon of fixed energy interacts with the electronic structure and removes an electron that escapes from the system with accurately measurable kinetic energy.
The difference between the energy of the primary photon and the kinetic energy of the emitted electron, provides the removal energy which coincides with the binding energy (BE) of the electron into the system.
XPS was established as one of the most powerful techniques to access the electronic structure and charged excitations.
Binding energies of electrons in occupied states, core or valence, can be measured with an accuracy of up to $10^{-3}$~eV by XPS.

From the theoretical point of view, the calculation of electron removal/addition energies is challenging \cite{FetterWalecka,MartinReiningCeperley}.
An exact analytic solution of the Schr\"odinger equation for electron BEs is only available for one-electron systems, e.g.\ the hydrogen atom.
Already in helium one should take into account an electron-electron interaction term in the Hamiltonian and faces a many-body problem \cite{LiOlevano17,LiOlevano19}.
The simplest and historically the first way to tackle this problem is by mean-field approaches, e.g.\ the Hartree or the Hartree-Fock   methods, in which the interaction of one electron with all other electrons is replaced by a mean-field potential self-consistently calculated.
In Hartree-Fock (HF) the Koopmans' theorem holds and states that HF eigenvalues are directly associated with electron removal/addition energies \cite{FetterWalecka,MartinReiningCeperley}.
However, the HF is an approximated method that neglects correlation energies.
On valence and core electron levels, this reflects in a systematic overestimation of BEs, as we will show.

Today, a more popular approach to tackle the many-body problem is density-functional theory (DFT) \cite{HohenbergKohn64,KohnSham65,JonesGunnarsson89,Martin}.
DFT is an in-principle exact approach to calculate the total ground-state energy.
In DFT electron removal/addition energies can be calculated exactly by the so-called $\Delta$SCF method \cite{JonesGunnarsson89,Martin} as total energy differences between the neutral and ionized systems.
The $\Delta$SCF method is exact only to calculate the highest occupied molecular orbital (HOMO) and the lowest unoccupied (LUMO) binding energies, that is the ionization potentials and the electron affinity.
It can be justified for the lowest levels within a given symmetry \cite{GunnarssonLundqvist76}.
For other levels, assumptions must be imposed on the relaxation of the ion, e.g.\ the localization of the core hole, leading to inaccuracy.
Furthermore, the $\Delta$SCF method can in principle be applied only to finite systems \cite{JonesGunnarsson89,Martin}.
In periodic solids some other assumptions/corrections, such as adding an infinite compensating charge on the background, are required, leading again to inaccuracies.
Nevertheless, even in approximated DFT, such as in local-density approximation (LDA) or beyond (PBE) \cite{LangrethMehl83,PerdewErnzerhof96}, the $\Delta$SCF method generally compares well with the experiment.
For the lightest elements deviations typically lie in the range 0.3$\sim$0.7~eV \cite{PueyoBellafontIllas16}, but can increase one order of magnitude \cite{GolzeRinke18}.

Always within DFT, Kohn-Sham (KS) eigenvalues are very often directly used to estimate electron removal/addition energies \cite{JonesGunnarsson89,Martin}.
This procedure is in principle exact only to evaluate the ionization potential equal to the last occupied KS energy \cite{PerdewBalduz82,LevySahni84,AlmbladhVonBarth85}.
Indeed, the Koopmans' theorem does not hold in DFT like in HF.
For all other states, this procedure is equivalent to look at DFT as a mean-field approximation, with the exchange-correlation (xc) potential $v_{xc}(r)$ as the mean-field.
To this error of principle, we must further add the error due to the unavoidable approximation on the xc functional of DFT.
As we will show, DFT KS energies in the PBE (or LDA) approximations systematically underestimate core and valence electron BEs, with an error that is larger than HF for finite systems.

In this work, we calculate electron binding energies of noble gas atoms within the framework of many-body perturbation theory (MBPT) and using the $GW$ approximation on the self-energy \cite{Hedin65,StrinatiHanke80,StrinatiHanke82,HybertsenLouie85,GodbySham87}.
MBPT or Green's function theory is an in-principle exact framework to calculate electron removal/addition energies which directly correspond to the poles of the one-particle Green's function \cite{FetterWalecka}.
The Green's function can be calculated via the Dyson equation from the self-energy \cite{FetterWalecka}, but the exact form of the latter is too complex for real systems.
Although in principle exact, MBPT must also resort to approximations.
The $GW$ approximation to the MBPT self-energy has demonstrated its validity on the band-gaps of solids \cite{Hedin99,MartinReiningCeperley}, and of the HOMO-LUMO gaps of molecules \cite{Li2019,Bruneval12,VanSettenRinke15}.
Here we benchmark $GW$ on core levels of atoms.
We choose in particular noble gas atoms as workbench since they are closed shell and electron energy levels are unaffected by other complications, like chemical shifts due to the molecular or solid-state environment.
To the best of our knowledge, this is the first time that the $GW$ approximation is tested at energies as deep as 100~keV.

\begin{table}[t]
\caption{Noble gas atoms (He to Xe) spin-orbit (SO) split in the ZORA and DK3 relativistic approximations on top of the DFT-PBE and HF approaches: mean absolute error (MAE in eV) and mean absolute percentage error (MAPE) with respect to the experiment \cite{Siegbahn,Lotz70}.}
\begin{tabular}{l|rrrr}
  & \multicolumn{2}{c}{MAE [eV]} & \multicolumn{2}{c}{MAPE} \\ 
SO (spin-orbit) split & ZORA &  DK3 &   ZORA &    DK3 \\
       \hline
DFT-PBE      & 0.35 & 17.6 & 3.1\% & 44.2\% \\
HF       & 1.18 & 17.1 & 10.0\% & 38.4\% \\
\end{tabular}
\label{SOerror}
\end{table}

\paragraph*{Methods. ---}
The starting point of our \textit{ab initio} procedure is a standard HF, or alternatively a DFT PBE calculation.
Relativistic effects are evaluated by the zero-order regular approximation (ZORA) \cite{ChangDurand86,HeullyMartensson-Pendrill86,vanLentheSnijders93} and also the third-order Douglas-Kroll (DK3) approximation \cite{NakajimaHirao00,NakajimaHirao00b}.
Their respective performances have been assessed with respect to experimental atomic spin-orbit splits which can be measured accurately without calibration problems.
They are resumed in Table~\ref{SOerror} which presents the mean absolute error (MAE in eV) and the mean absolute percentage error (MAPE) with respect to the experiment of the ZORA and the DK3 approximations on top of both DFT-PBE and HF (see the \appsuppl\ for detailed results).
In noble gas atoms the ZORA is more accurate than the DK3.
Unless differently specified, in the following we will only present and discuss ZORA results.
In any case, $GW$ corrections depend weakly from the relativistic approximation (see \appsuppl).

DFT KS or HF eigenvalues $E_i$ and eigenfunctions $\phi_i$ are then used to build the first-iteration Green's function,
\begin{equation}
  G(r,r',\omega) = \sum_i \frac{\phi(r) \phi^*_i(r')}{\omega - E_i - i\eta \sgn(\mu - E_i)}
  , \label{G}
\end{equation}
with $\mu$ the chemical potential and $\eta$ infinitesimal.
From $G$ we build the random-phase approximation (RPA) polarizability, $\Pi = -iGG$, and the screened Coulomb potential $W = w + w\Pi W$ (with $w = 1 / |r-r'|$ the bare Coulomb potential), and finally the self-energy in the $GW$ approximation,
\[
 \Sigma(r,r',\omega) = \frac{i}{2\pi} \int d\omega' \, e^{i\omega' \eta} G(r,r',\omega + \omega') W(r,r',\omega')
 ,
\]
where the $\omega'$ integral is carried on by contour deformation.
The $GW$ charged excitation quasiparticle energies are calculated by
\begin{equation}
 E^{GW}_i = E^\mathrm{H}_i + \langle \phi_i | \Sigma(\omega=E^{GW}_i) | \phi_i \rangle
 , \label{QP}
\end{equation}
where the $E^\mathrm{H}_i$ are the Hartree energies, i.e.\ the eigenvalues of the Hamiltonian only containing the kinetic, the nucleus external potential and the Hartree classical repulsion term.
The procedure can stop here to get the first iteration $G^0W^0$ energies, or a self-consistency can be carried on by recalculating $G$ in Eq.~(\ref{G}) with the new energies Eq.~(\ref{QP}).
All the calculation were performed using Gaussian basis sets.
We used an x2c-TZVPPall-2c \cite{Pollak_2017} segmented contracted Gaussian basis sets optimized at the one-electron exact two-component level, taking advantage of Coulomb-fitting resolution of the identity (RI-V) \cite{Duchemin_2017} with the auxiliary basis def2-universal-JKFIT \cite{Weigend_2008} for He-Kr and auxiliary basis generated by AutoAux \cite{Stoychev_2017} for Xe and Rn.
We used the codes \texttt{NWCHEM} \cite{nwchem} for the HF and DFT calculations, and  \texttt{Fiesta} \cite{BlaseOlevano11,Jac15a,Li16} with some checks by 
\texttt{TurboMole} \cite{turbomole} for $GW$.

\begin{figure}[t]
 \includegraphics[width=\columnwidth]{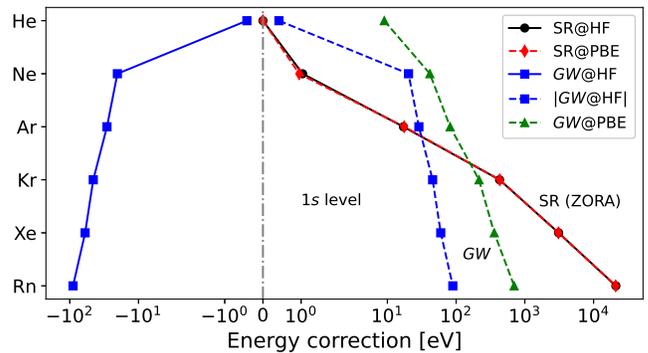}
 \caption{$GW$ vs ZORA scalar relativistic (SR) corrections to the PBE and HF energies of the 1$s$ level of noble gas atoms.}
 \label{GWvsSRZORA}
\end{figure}

\paragraph*{Results. ---}
Fig.~\ref{GWvsSRZORA} shows the relative magnitude of $GW$ many-body and ZORA scalar relativistic (SR) corrections to the HF and DFT-PBE energies of the 1$s$ level as a function of the atomic number $Z$.
ZORA scalar relativistic corrections do not depend on whether they are applied on top of PBE or HF, the two curves overlap (same for DK3, see \appsuppl).
They undergo a large increase with $Z$, going beyond 3~keV in Xe (20~keV in Rn).  
In contrast, $GW$ corrections on top of HF have small negative values, not going beyond $-100$~eV, so to reduce the slight overestimation of HF energies.
On the other hand, $GW$ corrections on top of PBE have a large increase with $Z$, so to reduce the large underestimation of PBE energies.
$GW$ corrections are larger than SR at small $Z$, and become smaller at large $Z$, although still not negligible (360~eV correction at Xe, 700~eV for Rn).
They can never be neglected.

\begin{figure}[t]
 \includegraphics[width=\columnwidth]{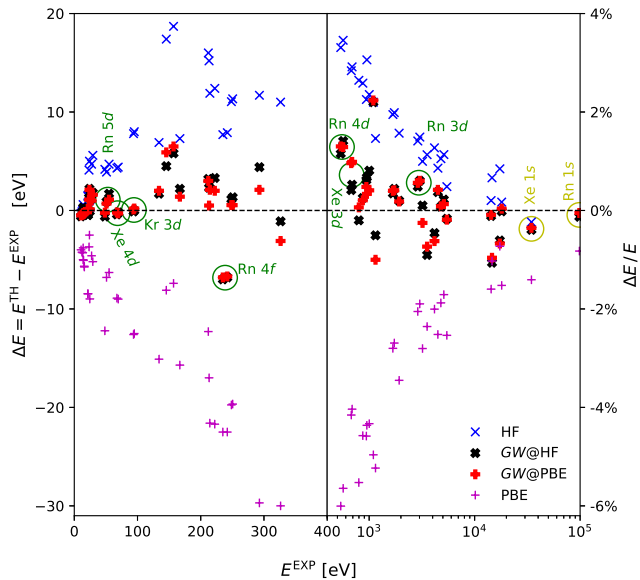}
 \caption{Noble gas atoms ZORA PBE, HF, and $GW$ electronic energy absolute (left, 0 to 400 eV) and relative (right, from 400 to $10^5$~eV) errors with respect to the experiment \cite{Siegbahn,Lotz70}.}
 \label{BEfig}
\end{figure}

Fig.~\ref{BEfig} shows the error with respect to the experiment on the electronic binding energies calculated within the ZORA relativistic scheme in all exchange-correlation approximations.
Full detailed results on the electronic binding energies can be found in the \appsuppl.
As experimental reference values we have taken the XPS binding energies reported in Ref.~\cite{Siegbahn}, except for Ne 2$p$, Ar 3$p$, Kr 1$s$ and 4$p$, Xe 1$s$ and 4$p_{1/2}$, and Rn which are taken from \cite{Lotz70}.
We estimate the former more direct and accurate, but values from the latter are not far when both are available.
In any case, our conclusions do not change if we use the similar values of Ref.~\cite{Carlson}, \cite{Williams} or \cite{BeardenBurr67} (see also Ref.~\cite{PiaSaracco11}).

As we already anticipated in the introduction, we can see from Fig.~\ref{BEfig} (see also \appsuppl) that DFT-PBE KS energies systematically underestimate experimental BEs.
The underestimation can be as large as 500 eV in Xe 1$s$ and 800~eV in Rn 1$s$.
However, since the Xe 1$s$ level is already 34500 eV deep, the relative error is only 1.4\%.
Starting from the deepest levels, the PBE underestimation relative error systematically increases, so to achieve almost 40\% in the shallowest levels.
This error precisely corresponds to the DFT-PBE (or LDA) systematical underestimation of the HOMO-LUMO gap for finite systems, and of the band-gap in infinite periodic solids.
On the other hand, in general, HF eigenvalues overestimate removal energies, though less systematically (for instance, Xe and Rn 1$s$ levels are underestimated).
However the HF error is smaller.
We can conclude that HF is better than DFT-PBE on noble gas atoms at all binding energies.

\begin{table}[t]
\caption{Noble gas atoms (He to Xe) electronic binding energies in the DFT-PBE, HF, $GW$ (on top of PBE and HF), ZORA and DK3 approximations: mean absolute error (MAE in eV) and mean absolute percentage error (MAPE) with respect to the experiment.}
\begin{tabular}{l|rrrr}
 & \multicolumn{2}{c}{MAE} & \multicolumn{2}{c}{MAPE} \\ 
BE (binding energies) & ZORA &  DK3 &   ZORA &    DK3 \\
       \hline
PBE      & 44.5 & 39.9 & 14.0\% & 14.1\% \\
HF       & 16.2 & 20.6 &  4.7\% &  4.8\% \\
$GW$@PBE &  6.0 & 10.4 &  1.2\% &  1.4\% \\
$GW$@HF  &  6.3 & 10.0 &  1.2\% &  1.4\% \\
\end{tabular}
\label{BEerror}
\end{table}

Always on Fig.~\ref{BEfig} we then immediately remark the net improvement brought by $GW$ calculations both on top of PBE and HF ($GW$@PBE and $GW$@HF respectively).
First note that our data are not the results of a single iteration $G^0W^0$ calculation, but we have performed a self-consistency on the quasiparticle energies only, whereas wavefunctions have been kept at their iteration 0 level, that is PBE or HF wavefunctions.
What is surprising is that, no matter the starting point, self-consistent $GW$ energies achieve almost the same value.
$GW$@PBE and $GW$@HF are distant only a few tenths of electronvolt on the shallowest energies and a few tenths of percentage on the deepest energies.
$GW$ self-consistency on energies only is then sufficient to get a result that is almost independent on the starting point.
And PBE and HF are among the most distant starting points, in practice, the two extremes in hybrid theories \cite{AtallaScheffler13}.
Our data indicate that PBE and HF wavefunctions are very close for noble gas atoms,  the main difference is in energies.

For a fair evaluation of the validity of the $GW$ approximation with respect to the experiment, we believe that a correct interpretation should take into account from one side the absolute error, $\Delta E = E^\mathrm{TH} - E^\mathrm{EXP}$, for the shallowest levels, as it is done in the left part from 0 to 400~eV of Fig.~\ref{BEfig}; and from another side, the relative error, $\Delta E / E^\mathrm{EXP}$, for the lowest-lying core levels which are placed thousands of electronvolt deep, as it is done in the right part of Fig.~\ref{BEfig} from 400~eV to 100 keV.
Indeed, the 0.1$\sim$0.2~eV accuracy achieved by $GW$ on low-energy valence and conduction levels is too pretentious in core electron binding energies whose magnitude can be 5 orders larger.
With this key to understanding, the results we obtained on noble gas atoms by the $GW$ approximation are in very good agreement with the experiment.
Indeed, $GW$ errors on the shallowest valence electrons are always within the few tenths of electronvolt, as usually found for $GW$ in both chemistry and solid-state physics.
At the same time, $GW$ errors are always below 1\% in deep core levels.
The improvement from HF and PBE is quantified in Table~\ref{BEerror} which presents statistical averages over all energies from He to Xe.
Both the MAE and the MAPE are strongly reduced when passing from either PBE or HF to $GW$.
Most importantly, the $GW$ self-energy contains the right and valid physics since it is able to both reduce the PBE underestimation and also the HF overestimation, both directions. 
The present results represent \textit{a surprising confirmation of the $GW$ approximation whose validity is thus certified even at high energies}, tens of keV.

A last but not least point to be remarked is the surprising accuracy of $GW$ on $d$ electrons, but also on the $f$.
We first notice that, among all levels, $d$ and $f$ electrons are those where the $GW$@PBE and $GW$@HF values are the closest in energy, indicating that the PBE and HF wavefunctions are very close.
This is very surprising for levels where exchange and correlation are supposed to play a major role.
The $GW$ quasiparticle renormalization factor is $Z=0.87 \pm 0.03$ on the full set of $d$ and $f$ electrons, except Xe 4$d$ where $Z=0.61$.
Furthermore, the shallowest $d$ electrons (Kr 3$d$, Xe 4$d$ and Rn 5$d$, see Fig.~\ref{BEfig}) are also the levels where $GW$ achieves one of the best agreements with the experiment in absolute values.
Whereas on the deepest the relative error is at 0.5\% for Rn (and also Xe) 3$d$ and raises to 1.3\% in Rn 4$d$.
On Rn 4$f$ electrons $GW$ correlations correct the HF 3.3\% overestimation, but with an overshot, at the end achieving a -2.9\% underestimation.
We can conclude that \textit{$GW$ describes quite well $d$-electron correlations and slightly overestimates $f$-electron correlation energy}.

In general, the largest errors are registered at the level of the outermost $s$ electrons (e.g.\ Rn 6$s$, Xe 5$s$ or Kr 4$s$) due to difficult convergence in the $GW$ iterations.
Another source of discrepancy is the fact that our $GW$ is not fully self-consistent and it is self-consistent only on energies.
The order of this error can be estimated from the difference between $GW$@PBE and $GW$@HF energies which presents also some variability along with the table.
Smaller than this is the error due to the cutoff on $GW$ parameters (number of unoccupied states in the calculation for $W$ and $\Sigma$) and on the basis set.
The order of the error due to the relativistic approximation can be estimated by comparing the two relativistic approaches considered here, ZORA and DK3 (see \appsuppl).
However, all Pauli spinor lowest order $v/c$ relativistic developments are expected to break down at large $Z$ where also anti-matter negative energies enter into play and one should solve the full relativistic Dirac equation.
Furthermore, beyond the here considered single-particle, \textit{many-particle} relativistic effects should be taken into account, e.g.\ the Breit interaction, spin-of-one-electron orbit-of-another-electron, orbit-orbit, spin-spin etc. \cite{OlevanoLadisa10}.
These effects are very difficult to include but should be of the same order as the single-particle spin-orbit.
Nuclear finite mass effects, i.e.\ the reduced mass of electrons and the mass polarization term, are also here neglected but well present in the experiment, although the recoil energy of the final ion out of the XPS experiment has already been removed to provide corrected binding energies \cite{Siegbahn}.
Finally, the experiment also contains quantum electrodynamics (QED) radiative corrections, but these do not grow with $Z$.
The discrepancy due to the $GW$ approximation itself, that is the neglect of vertex corrections into the many-body self-energy and into the polarizability, is the residual once eliminated all previous sources of error.

\paragraph*{Conclusions. ---} 
We workbenched the $GW$ approximation at high energies ($\sim$10$^5$~eV) with respect to core and valence electron removal energies.
$GW$ is in very good agreement with XPS binding energies, with a mean relative error of 1.2\%.
The largest discrepancies are observed at the level of the outermost $s$ levels, whereas correlations in $d$ and even in $f$ electrons are surprisingly well described by $GW$.

\paragraph*{Acknowledgments. ---}
We thank X. Blase and I. Duchemin for useful discussions.
Part of the calculations were using the allocation of computational resources from GENCI–IDRIS (Grant 2021-A0110912036).

\appendix

\subsection*{APPENDIX: Full detailed results}
\paragraph*{ZORA vs DK3 relativistic corrections. ---}
In Table~\ref{SO} we present the spin-orbit split of $p$ and $d$ electronic levels as provided by the ZORA and DK3 relativistic approximations in comparison to the experiment.
In noble gas atoms, both the mean absolute error (MAE) and the mean absolute percentage error (MAPE) are smaller in ZORA than in DK3.

\begin{table}[h!]
\begin{tabular}{llD{.}{.}{3}D{.}{.}{3}D{.}{.}{3}D{.}{.}{3}D{.}{.}{3}}
\multicolumn{2}{l}{SO split} & \multicolumn{2}{c}{DK3} & \multicolumn{2}{c}{ZORA}  \\
Atom & Orbit & \multicolumn{1}{c}{\textrm{PBE}} & \multicolumn{1}{c}{\textrm{HF}} & \multicolumn{1}{c}{\textrm{PBE}} & \multicolumn{1}{c}{\textrm{HF}} & \multicolumn{1}{c}{\textrm{EXP}}\\ 
\hline
Ne & 2$p$ & 0.06 & 0.08 & 0.10 & 0.13 & 0.10\\
Ar & 2$p$ & 1.22 & 1.32 & 2.17 & 2.34 & 2.11\\
 & 3$p$ & 0.10 & 0.12 & 0.17 & 0.21 & 0.18\\
Kr & 2$p$ & 27.9 & 28.6 & 52.9 & 54.3 & 52.5\\
 & 3$p$ & 4.1 & 4.4 & 7.8 & 8.4 & 7.8\\
 & 3$d$ & 0.9 & 0.9 & 1.3 & 1.4 & 1.2\\
 & 4$p$ & 0.33 & 0.39 & 0.62 & 0.73 & 0.67\\
Xe & 2$p$ & 163.1 & 165.8 & 322.5 & 327.6 & 319.9\\
 & 3$p$ & 31.1 & 32.4 & 61.2 & 63.9 & 61.5\\
 & 3$d$ & 7.8 & 8.1 & 12.9 & 13.4 & 12.6\\
 & 4$p$ & 6.2 & 6.5 & 12.2 & 12.8 & 11.5\\
 & 4$d$ & 1.2 & 1.3 & 2.0 & 2.1 & 2.0\\
 & 5$p$ & 0.62 & 0.73 & 1.22 & 1.43 & 1.27\\
 \hline
 MAE & & 17.6 & 17.1 & 0.35 & 1.18 & \\
 MAPE &  & 44.2\% & 38.4\% & 3.1\% & 10.0\% & 
\end{tabular}
\caption{Spin-orbit split in eV as provided by the DK3 and ZORA approaches in both DFT-PBE and HF compared to the experiment (XPS of Ref.~ \cite{Siegbahn}, except Ne 2$p$, Ar 3$p$, Kr 4$p$, Xe 4$p_{1/2}$ which are taken from \cite{Lotz70}).
The last lines presents the mean absolute error (MAE) and the mean absolute percentage error (MAPE) with respect to the experiment.}
\label{SO}
\end{table}

\begin{figure}[ht]
 \includegraphics[width=\columnwidth]{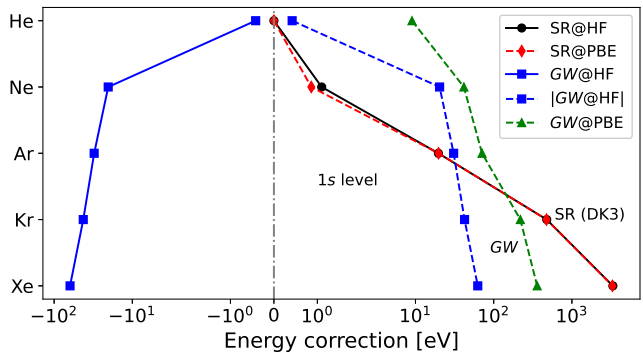}
 \caption{$GW$ vs DK3 scalar relativistic (SR) corrections to the PBE and HF energies of the 1$s$ level of noble gas atoms.}
 \label{GWvsSRDK3}
\end{figure}

\paragraph*{DK3 scalar relativistic vs $GW$ corrections. ---}
In Fig.~\ref{GWvsSRDK3} we present DK3 scalar relativistic (SR) with respect to $GW$ corrections for the 1$s$ level of He to Xe, which is similar to the figure in main text for ZORA scalar relativistic.

\paragraph*{Full ZORA and DK3 $GW$ results. ---}
In the main text we have discussed almost exclusively the ZORA results which are better in agreement with the experiment both on the spin-orbit split and on the absolute position of levels provided by the scalar relativistic corrections.
Here we provide the full results for binding energies and errors for both the ZORA and the DK3 relativistic schemes, separately in Table~\ref{BE} and \ref{DK3} respectively.
As experimental reference values we have taken the XPS binding energies reported in Ref.~\cite{Siegbahn}, except for Ne 2$p$, Ar 3$p$, Kr 1$s$ and 4$p$, Xe 1$s$ and 4$p_{1/2}$, and Rn which are taken from \cite{Lotz70}.
We estimate the former more direct and (may be) accurate, but values from the latter are not far, when both are available.
In any case, our conclusions do not change if we use the similar values of Ref.~\cite{Carlson}, \cite{Williams} or \cite{BeardenBurr67} (see also Ref.~\cite{PiaSaracco11}).

\begin{table*}[p]
\footnotesize
\begin{tabular}{ll|rrr|rrr|D{.}{.}{2}|rrr|rrr}
 \multicolumn{2}{l|}{ZORA} & \multicolumn{3}{c|}{PBE} & \multicolumn{3}{c|}{$GW$@PBE} & \multicolumn{1}{c|}{\textrm{EXP}} &  \multicolumn{3}{c|}{$GW$@HF} & \multicolumn{3}{c}{HF}  \\
Atom & Orbital & $E$ & $\Delta E$ & $\Delta E/E$ & $E$ & $\Delta E$ & $\Delta E/E$ &  \multicolumn{1}{c|}{$E$} & $E$ & $\Delta E$ & $\Delta E/E$ & $E$ & $\Delta E$ & $\Delta E/E$ \\
\hline
He & 1$s$ & 15.6 & -9.0 & -36.4\% & 24.7 & 0.1 & 0.3\% & 24.59 & 24.5 & -0.1 & -0.2\% & 24.9 & 0.4 & 1.5\% \\
Ne & 1$s$ & 830.4 & -39.8 & -4.6\% & 872.0 & 1.8 & 0.2\% & 870.2 & 872.4 & 2.2 & 0.3\% & 892.7 & 22.5 & 2.6\% \\
 & 2$s$ & 36.2 & -12.2 & -25.3\% & 48.2 & -0.2 & -0.5\% & 48.42 & 47.8 & -0.6 & -1.3\% & 52.6 & 4.2 & 8.6\% \\
 & 2$p_{1/2}$ & 13.2 & -8.5 & -39.1\% & 21.8 & 0.1 & 0.4\% & 21.66 & 21.2 & -0.4 & -1.9\% & 23.2 & 1.5 & 7.0\% \\
 & 2$p_{3/2}$ & 13.1 & -8.5 & -39.2\% & 21.7 & 0.1 & 0.4\% & 21.56 & 21.1 & -0.4 & -2.1\% & 23.0 & 1.5 & 6.9\% \\
Ar & 1$s$ & 3116.0 & -89.9 & -2.8\% & 3197.8 & -8.1 & -0.3\% & 3205.9 & 3209.1 & 3.2 & 0.1\% & 3237.9 & 31.9 & 1.0\% \\
 & 2$s$ & 296.3 & -30.0 & -9.2\% & 323.2 & -3.1 & -0.9\% & 326.3 & 325.2 & -1.1 & -0.3\% & 337.3 & 11.0 & 3.4\% \\
 & 2$p_{1/2}$ & 230.9 & -19.7 & -7.8\% & 251.1 & 0.6 & 0.2\% & 250.56 & 251.9 & 1.4 & 0.5\% & 261.9 & 11.3 & 4.5\% \\
 & 2$p_{3/2}$ & 228.7 & -19.7 & -7.9\% & 249.0 & 0.5 & 0.2\% & 248.45 & 249.6 & 1.1 & 0.5\% & 259.5 & 11.1 & 4.5\% \\
 & 3$s$ & 24.1 & -5.2 & -17.7\% & 30.9 & 1.6 & 5.5\% & 29.3 & 31.1 & 1.8 & 6.1\% & 34.9 & 5.6 & 19.3\% \\
 & 3$p_{1/2}$ & 10.2 & -5.7 & -35.8\% & 15.7 & -0.3 & -1.6\% & 15.94 & 15.8 & -0.1 & -0.9\% & 16.2 & 0.2 & 1.4\% \\
 & 3$p_{3/2}$ & 10.1 & -5.7 & -36.1\% & 15.5 & -0.2 & -1.5\% & 15.76 & 15.6 & -0.2 & -1.1\% & 16.0 & 0.2 & 1.2\% \\
Kr & 1$s$ & 14098.2 & -228.8 & -1.6\% & 14314.3 & -12.7 & -0.1\% & 14327 & 14310.9 & -16.1 & -0.1\% & 14355.9 & 28.9 & 0.2\% \\
 & 2$s$ & 1858.2 & -66.4 & -3.4\% & 1928.1 & 3.5 & 0.2\% & 1924.6 & 1928.4 & 3.8 & 0.2\% & 1954.8 & 30.2 & 1.6\% \\
 & 2$p_{1/2}$ & 1684.3 & -46.6 & -2.7\% & 1738.0 & 7.1 & 0.4\% & 1730.9 & 1738.5 & 7.6 & 0.4\% & 1765.3 & 34.4 & 2.0\% \\
 & 2$p_{3/2}$ & 1631.4 & -47.0 & -2.8\% & 1685.1 & 6.7 & 0.4\% & 1678.4 & 1684.2 & 5.8 & 0.3\% & 1711.1 & 32.7 & 1.9\% \\
 & 3$s$ & 263.1 & -29.7 & -10.1\% & 294.9 & 2.1 & 0.7\% & 292.8 & 297.2 & 4.4 & 1.5\% & 304.5 & 11.7 & 4.0\% \\
 & 3$p_{1/2}$ & 200.5 & -21.7 & -9.8\% & 224.2 & 2.0 & 0.9\% & 222.2 & 225.5 & 3.3 & 1.5\% & 234.6 & 12.4 & 5.6\% \\
 & 3$p_{3/2}$ & 192.8 & -21.6 & -10.1\% & 216.5 & 2.1 & 1.0\% & 214.4 & 217.1 & 2.7 & 1.3\% & 226.3 & 11.9 & 5.5\% \\
 & 3$d_{3/2}$ & 82.4 & -12.5 & -13.2\% & 95.1 & 0.2 & 0.2\% & 94.9 & 94.9 & 0.0 & 0.0\% & 102.9 & 8.0 & 8.4\% \\
 & 3$d_{5/2}$ & 81.1 & -12.6 & -13.4\% & 93.8 & 0.1 & 0.1\% & 93.7 & 93.6 & -0.1 & -0.2\% & 101.5 & 7.8 & 8.4\% \\
 & 4$s$ & 22.8 & -4.6 & -16.7\% & 28.4 & 1.0 & 3.7\% & 27.4 & 28.6 & 1.2 & 4.5\% & 32.1 & 4.7 & 17.3\% \\
 & 4$p_{1/2}$ & 9.6 & -5.0 & -34.4\% & 14.3 & -0.4 & -2.5\% & 14.67 & 14.4 & -0.3 & -1.7\% & 14.7 & 0.1 & 0.4\% \\
 & 4$p_{3/2}$ & 9.0 & -5.0 & -35.7\% & 13.7 & -0.3 & -2.3\% & 14 & 13.7 & -0.3 & -2.2\% & 14.0 & 0.0 & -0.1\% \\
Xe & 1$s$ & 34078.5 & -486.5 & -1.4\% & 34443.0 & -122.0 & -0.4\% & 34565 & 34429.0 & -136.0 & -0.4\% & 34483.5 & -81.5 & -0.2\% \\
 & 2$s$ & 5314.9 & -138.3 & -2.5\% & 5443.0 & -10.2 & -0.2\% & 5453.2 & 5444.0 & -9.2 & -0.2\% & 5479.6 & 26.4 & 0.5\% \\
 & 2$p_{1/2}$ & 5019.9 & -87.3 & -1.7\% & 5113.8 & 6.6 & 0.1\% & 5107.2 & 5118.5 & 11.3 & 0.2\% & 5165.2 & 58.0 & 1.1\% \\
 & 2$p_{3/2}$ & 4697.4 & -89.9 & -1.9\% & 4791.4 & 4.1 & 0.1\% & 4787.3 & 4790.9 & 3.6 & 0.1\% & 4837.6 & 50.3 & 1.1\% \\
 & 3$s$ & 1088.6 & -60.1 & -5.2\% & 1137.2 & -11.5 & -1.0\% & 1148.7 & 1142.9 & -5.8 & -0.5\% & 1165.5 & 16.8 & 1.5\% \\
 & 3$p_{1/2}$ & 958.7 & -43.4 & -4.3\% & 1006.2 & 4.1 & 0.4\% & 1002.1 & 1010.2 & 8.1 & 0.8\% & 1025.7 & 23.6 & 2.4\% \\
 & 3$p_{3/2}$ & 897.5 & -43.1 & -4.6\% & 945.1 & 4.5 & 0.5\% & 940.6 & 946.4 & 5.8 & 0.6\% & 961.8 & 21.2 & 2.3\% \\
 & 3$d_{3/2}$ & 661.2 & -27.8 & -4.0\% & 695.8 & 6.8 & 1.0\% & 689 & 692.6 & 3.6 & 0.5\% & 709.1 & 20.1 & 2.9\% \\
 & 3$d_{5/2}$ & 648.3 & -28.1 & -4.2\% & 682.9 & 6.5 & 1.0\% & 676.4 & 679.2 & 2.8 & 0.4\% & 695.6 & 19.2 & 2.8\% \\
 & 4$s$ & 196.2 & -17.0 & -8.0\% & 213.7 & 0.5 & 0.2\% & 213.2 & 215.5 & 2.3 & 1.1\% & 228.4 & 15.2 & 7.1\% \\
 & 4$p_{1/2}$ & 149.6 & -7.4 & -4.7\% & 163.5 & 6.5 & 4.2\% & 157 & 162.8 & 5.8 & 3.7\% & 175.7 & 18.7 & 11.9\% \\
 & 4$p_{3/2}$ & 137.4 & -8.1 & -5.5\% & 151.4 & 5.9 & 4.0\% & 145.5 & 150.0 & 4.5 & 3.1\% & 162.9 & 17.4 & 11.9\% \\
 & 4$d_{3/2}$ & 60.5 & -9.0 & -12.9\% & 69.2 & -0.3 & -0.4\% & 69.5 & 69.3 & -0.2 & -0.3\% & 73.9 & 4.4 & 6.3\% \\
 & 4$d_{5/2}$ & 58.6 & -8.9 & -13.2\% & 67.3 & -0.2 & -0.4\% & 67.5 & 67.1 & -0.4 & -0.5\% & 71.8 & 4.3 & 6.3\% \\
 & 5$s$ & 19.6 & -3.7 & -16.0\% & 24.4 & 1.1 & 4.7\% & 23.3 & 24.6 & 1.3 & 5.7\% & 27.4 & 4.1 & 17.6\% \\
 & 5$p_{1/2}$ & 9.1 & -4.3 & -32.3\% & 13.0 & -0.4 & -3.2\% & 13.4 & 13.3 & -0.1 & -0.7\% & 13.4 & 0.0 & 0.3\% \\
 & 5$p_{3/2}$ & 7.9 & -4.3 & -35.3\% & 11.7 & -0.4 & -3.2\% & 12.13 & 11.9 & -0.3 & -2.1\% & 12.0 & -0.1 & -1.0\% \\
Rn & 1$s$ & 97593.1 & -810.9 & -0.8\% & 98353.6 & -50.4 & -0.1\% & 98404 & 98276.9 & -127.1 & -0.1\% & 98308.1 & -95.9 & -0.1\% \\
 & 2$s$ & 17780.2 & -274.8 & -1.5\% & 18061.5 & 6.5 & 0.0\% & 18055 & 18051.8 & -3.2 & 0.0\% & 18085.7 & 30.7 & 0.2\% \\
 & 2$p_{1/2}$ & 17207.9 & -126.1 & -0.7\% & 17217.7 & -116.3 & -0.7\% & 17334 & 17228.4 & -105.6 & -0.6\% & 17480.7 & 146.7 & 0.8\% \\
 & 2$p_{3/2}$ & 14464.8 & -150.2 & -1.0\% & 14474.6 & -140.4 & -1.0\% & 14615 & 14459.8 & -155.2 & -1.1\% & 14712.0 & 97.0 & 0.7\% \\
 & 3$s$ & 4370.1 & -112.9 & -2.5\% & 4501.8 & 18.8 & 0.4\% & 4483 & 4499.7 & 16.7 & 0.4\% & 4521.5 & 38.5 & 0.9\% \\
 & 3$p_{1/2}$ & 4078.6 & -83.4 & -2.0\% & 4136.1 & -25.9 & -0.6\% & 4162 & 4142.9 & -19.1 & -0.5\% & 4215.0 & 53.0 & 1.3\% \\
 & 3$p_{3/2}$ & 3458.5 & -83.5 & -2.4\% & 3516.0 & -26.0 & -0.7\% & 3542 & 3510.0 & -32.0 & -0.9\% & 3582.1 & 40.1 & 1.1\% \\
 & 3$d_{3/2}$ & 2961.7 & -57.3 & -1.9\% & 3037.2 & 18.2 & 0.6\% & 3019 & 3037.2 & 18.2 & 0.6\% & 3064.0 & 45.0 & 1.5\% \\
 & 3$d_{5/2}$ & 2830.7 & -59.3 & -2.1\% & 2906.2 & 16.2 & 0.6\% & 2890 & 2904.0 & 14.0 & 0.5\% & 2930.8 & 40.8 & 1.4\% \\
 & 4$s$ & 1041.6 & -54.4 & -5.0\% & 1120.5 & 24.5 & 2.2\% & 1096 & 1120.1 & 24.1 & 2.2\% & 1120.1 & 24.1 & 2.2\% \\
 & 4$p_{1/2}$ & 909.5 & -41.5 & -4.4\% & 954.2 & 3.2 & 0.3\% & 951 & 957.4 & 6.4 & 0.7\% & 980.1 & 29.1 & 3.1\% \\
 & 4$p_{3/2}$ & 753.9 & -44.1 & -5.5\% & 798.5 & 0.5 & 0.1\% & 798 & 796.4 & -1.6 & -0.2\% & 819.1 & 21.1 & 2.6\% \\
 & 4$d_{3/2}$ & 535.0 & -32.0 & -5.6\% & 574.3 & 7.3 & 1.3\% & 567 & 575.0 & 8.0 & 1.4\% & 586.6 & 19.6 & 3.5\% \\
 & 4$d_{5/2}$ & 505.7 & -32.3 & -6.0\% & 545.0 & 7.0 & 1.3\% & 538 & 544.2 & 6.2 & 1.1\% & 555.8 & 17.8 & 3.3\% \\
 & 4$f_{5/2}$ & 219.5 & -22.5 & -9.3\% & 235.3 & -6.7 & -2.8\% & 242 & 235.2 & -6.8 & -2.8\% & 249.9 & 7.9 & 3.2\% \\
 & 4$f_{7/2}$ & 212.5 & -22.5 & -9.6\% & 228.2 & -6.8 & -2.9\% & 235 & 228.0 & -7.0 & -3.0\% & 242.7 & 7.7 & 3.3\% \\
 & 5$s$ & 199.7 & -12.3 & -5.8\% & 215.0 & 3.0 & 1.4\% & 212 & 215.2 & 3.2 & 1.5\% & 228.0 & 16.0 & 7.6\% \\
 & 5$p_{1/2}$ & 151.3 & -15.7 & -9.4\% & 168.4 & 1.4 & 0.8\% & 167 & 169.2 & 2.2 & 1.3\% & 174.3 & 7.3 & 4.3\% \\
 & 5$p_{3/2}$ & 118.9 & -15.1 & -11.3\% & 136.0 & 2.0 & 1.5\% & 134 & 135.7 & 1.7 & 1.3\% & 140.9 & 6.9 & 5.1\% \\
 & 5$d_{3/2}$ & 48.7 & -6.3 & -11.5\% & 56.1 & 1.1 & 2.1\% & 55 & 56.7 & 1.7 & 3.1\% & 59.7 & 4.7 & 8.6\% \\
 & 5$d_{5/2}$ & 44.2 & -6.8 & -13.3\% & 51.7 & 0.7 & 1.3\% & 51 & 51.9 & 0.9 & 1.7\% & 54.9 & 3.9 & 7.6\% \\
 & 6$s$ & 21.5 & -2.5 & -10.3\% & 26.1 & 2.1 & 8.7\% & 24 & 26.2 & 2.2 & 9.1\% & 29.0 & 5.0 & 20.8\% \\
 & 6$p_{1/2}$ & 10.2 & -3.8 & -26.9\% & 13.7 & -0.3 & -2.1\% & 14 & 14.3 & 0.3 & 2.2\% & 14.6 & 0.6 & 4.4\% \\
 & 6$p_{3/2}$ & 6.7 & -4.0 & -37.0\% & 10.2 & -0.5 & -4.6\% & 10.7 & 10.1 & -0.6 & -5.6\% & 10.4 & -0.3 & -2.6\% \\
\end{tabular}
\caption{ZORA electron binding energies (eV) with their absolute and relative errors with respect to the experiment.
}
\label{BE}
\end{table*}

\begin{table*}[t!]
\footnotesize
\begin{tabular}{ll|rrr|rrr|D{.}{.}{2}|rrr|rrr}
 \multicolumn{2}{l|}{DK3} & \multicolumn{3}{c|}{PBE} & \multicolumn{3}{c|}{$GW$@PBE} & \multicolumn{1}{c|}{\textrm{EXP}} &  \multicolumn{3}{c|}{$GW$@HF} & \multicolumn{3}{c}{HF}  \\
Atom & Orbital & $E$ & $\Delta E$ & $\Delta E/E$ & $E$ & $\Delta E$ & $\Delta E/E$ &  \multicolumn{1}{c|}{$E$} & $E$ & $\Delta E$ & $\Delta E/E$ & $E$ & $\Delta E$ & $\Delta E/E$ \\
\hline
He & 1$s$ & 15.6 & -9.0 & -36.4\% & 24.7 & 0.1 & 0.3\% & 24.59 & 24.5 & -0.1 & -0.2\% & 24.9 & 0.4 & 1.5\% \\
Ne & 1$s$ & 830.4 & -39.8 & -4.6\% & 871.9 & 1.7 & 0.2\% & 870.2 & 872.5 & 2.3 & 0.3\% & 892.8 & 22.6 & 2.6\% \\
 & 2$s$ & 36.2 & -12.2 & -25.2\% & 48.2 & -0.2 & -0.4\% & 48.42 & 47.8 & -0.6 & -1.3\% & 52.6 & 4.2 & 8.7\% \\
 & 2$p_{1/2}$ & 13.2 & -8.5 & -39.2\% & 21.7 & 0.1 & 0.3\% & 21.66 & 21.2 & -0.5 & -2.1\% & 23.1 & 1.5 & 6.8\% \\
 & 2$p_{3/2}$ & 13.1 & -8.4 & -39.2\% & 21.7 & 0.1 & 0.5\% & 21.56 & 21.1 & -0.4 & -2.0\% & 23.1 & 1.5 & 7.0\% \\
Ar & 1$s$ & 3118.1 & -87.8 & -2.7\% & 3189.0 & -16.9 & -0.5\% & 3205.9 & 3209.6 & 3.7 & 0.1\% & 3240.4 & 34.5 & 1.1\% \\
 & 2$s$ & 296.7 & -29.6 & -9.1\% & 323.4 & -2.9 & -0.9\% & 326.3 & 325.4 & -0.9 & -0.3\% & 337.6 & 11.3 & 3.5\% \\
 & 2$p_{1/2}$ & 230.4 & -20.2 & -8.0\% & 250.6 & 0.1 & 0.0\% & 250.56 & 251.4 & 0.8 & 0.3\% & 261.3 & 10.8 & 4.3\% \\
 & 2$p_{3/2}$ & 229.2 & -19.3 & -7.8\% & 249.4 & 1.0 & 0.4\% & 248.45 & 250.1 & 1.6 & 0.7\% & 260.0 & 11.6 & 4.7\% \\
 & 3$s$ & 24.1 & -5.2 & -17.7\% & 30.9 & 1.6 & 5.5\% & 29.3 & 31.1 & 1.8 & 6.1\% & 35.0 & 5.7 & 19.3\% \\
 & 3$p_{1/2}$ & 10.2 & -5.8 & -36.1\% & 15.6 & -0.3 & -1.9\% & 15.94 & 15.7 & -0.2 & -1.3\% & 16.1 & 0.2 & 1.0\% \\
 & 3$p_{3/2}$ & 10.1 & -5.7 & -36.0\% & 15.5 & -0.2 & -1.4\% & 15.76 & 15.6 & -0.1 & -0.9\% & 16.0 & 0.2 & 1.4\% \\
Kr & 1$s$ & 14140.9 & -186.1 & -1.3\% & 14360.2 & 33.1 & 0.2\% & 14327 & 14361.1 & 34.1 & 0.2\% & 14403.4 & 76.4 & 0.5\% \\
 & 2$s$ & 1863.4 & -61.2 & -3.2\% & 1930.7 & 6.1 & 0.3\% & 1924.6 & 1932.4 & 7.8 & 0.4\% & 1960.5 & 35.9 & 1.9\% \\
 & 2$p_{1/2}$ & 1666.3 & -64.6 & -3.7\% & 1720.4 & -10.5 & -0.6\% & 1730.9 & 1720.7 & -10.2 & -0.6\% & 1747.2 & 16.3 & 0.9\% \\
 & 2$p_{3/2}$ & 1638.4 & -40.0 & -2.4\% & 1692.5 & 14.1 & 0.8\% & 1678.4 & 1692.1 & 13.7 & 0.8\% & 1718.6 & 40.2 & 2.4\% \\
 & 3$s$ & 263.8 & -29.0 & -9.9\% & 295.1 & 2.3 & 0.8\% & 292.8 & 297.9 & 5.1 & 1.7\% & 305.3 & 12.5 & 4.3\% \\
 & 3$p_{1/2}$ & 197.9 & -24.3 & -11.0\% & 221.6 & -0.6 & -0.3\% & 222.2 & 222.7 & 0.5 & 0.2\% & 231.8 & 9.6 & 4.3\% \\
 & 3$p_{3/2}$ & 193.7 & -20.7 & -9.6\% & 217.5 & 3.1 & 1.4\% & 214.4 & 218.2 & 3.8 & 1.8\% & 227.4 & 13.0 & 6.0\% \\
 & 3$d_{3/2}$ & 82.0 & -12.9 & -13.6\% & 94.7 & -0.2 & -0.2\% & 94.9 & 94.5 & -0.4 & -0.4\% & 102.5 & 7.6 & 8.0\% \\
 & 3$d_{5/2}$ & 81.1 & -12.6 & -13.4\% & 93.8 & 0.1 & 0.1\% & 93.7 & 93.6 & -0.1 & -0.2\% & 101.6 & 7.9 & 8.4\% \\
 & 4$s$ & 23.0 & -4.4 & -16.2\% & 28.5 & 1.1 & 4.2\% & 27.4 & 28.7 & 1.3 & 4.9\% & 32.3 & 4.9 & 17.8\% \\
 & 4$p_{1/2}$ & 9.4 & -5.2 & -35.8\% & 14.1 & -0.6 & -3.8\% & 14.67 & 14.2 & -0.5 & -3.3\% & 14.5 & -0.2 & -1.3\% \\
 & 4$p_{3/2}$ & 9.1 & -4.9 & -35.1\% & 13.8 & -0.2 & -1.6\% & 14 & 13.8 & -0.2 & -1.5\% & 14.1 & 0.1 & 0.6\% \\
Xe & 1$s$ & 34298.6 & -266.4 & -0.8\% & 34659.2 & 94.2 & 0.3\% & 34565 & 34662.7 & 97.7 & 0.3\% & 34724.2 & 159.2 & 0.5\% \\
 & 2$s$ & 5338.6 & -114.6 & -2.1\% & 5470.6 & 17.4 & 0.3\% & 5453.2 & 5468.6 & 15.4 & 0.3\% & 5506.4 & 53.2 & 1.0\% \\
 & 2$p_{1/2}$ & 4901.1 & -206.1 & -4.0\% & 5001.4 & -105.8 & -2.1\% & 5107.2 & 5005.8 & -101.4 & -2.0\% & 5046.0 & -61.2 & -1.2\% \\
 & 2$p_{3/2}$ & 4738.0 & -49.3 & -1.0\% & 4838.3 & 51.0 & 1.1\% & 4787.3 & 4840.0 & 52.7 & 1.1\% & 4880.2 & 92.9 & 1.9\% \\
 & 3$s$ & 1092.2 & -56.5 & -4.9\% & 1136.5 & -12.2 & -1.1\% & 1148.7 & 1144.2 & -4.5 & -0.4\% & 1169.9 & 21.2 & 1.8\% \\
 & 3$p_{1/2}$ & 936.4 & -65.7 & -6.6\% & 985.1 & -17.0 & -1.7\% & 1002.1 & 988.3 & -13.8 & -1.4\% & 1002.6 & 0.5 & 0.0\% \\
 & 3$p_{3/2}$ & 905.3 & -35.3 & -3.8\% & 954.0 & 13.4 & 1.4\% & 940.6 & 955.9 & 15.3 & 1.6\% & 970.2 & 29.6 & 3.1\% \\
 & 3$d_{3/2}$ & 657.4 & -31.6 & -4.6\% & 686.7 & -2.3 & -0.3\% & 689 & 688.4 & -0.6 & -0.1\% & 705.2 & 16.2 & 2.3\% \\
 & 3$d_{5/2}$ & 649.5 & -26.9 & -4.0\% & 678.9 & 2.5 & 0.4\% & 676.4 & 680.3 & 3.9 & 0.6\% & 697.0 & 20.6 & 3.0\% \\
 & 4$s$ & 197.1 & -16.1 & -7.5\% & 214.4 & 1.2 & 0.6\% & 213.2 & 216.2 & 3.0 & 1.4\% & 229.4 & 16.2 & 7.6\% \\
 & 4$p_{1/2}$ & 145.3 & -11.7 & -7.4\% & 159.5 & 2.5 & 1.6\% & 157 & 158.5 & 1.5 & 1.0\% & 171.2 & 14.2 & 9.1\% \\
 & 4$p_{3/2}$ & 139.1 & -6.4 & -4.4\% & 153.3 & 7.8 & 5.4\% & 145.5 & 152.0 & 6.5 & 4.5\% & 164.7 & 19.2 & 13.2\% \\
 & 4$d_{3/2}$ & 60.1 & -9.4 & -13.6\% & 68.8 & -0.7 & -1.1\% & 69.5 & 68.7 & -0.8 & -1.1\% & 73.4 & 3.9 & 5.6\% \\
 & 4$d_{5/2}$ & 58.9 & -8.6 & -12.8\% & 67.5 & 0.0 & 0.1\% & 67.5 & 67.4 & -0.1 & -0.2\% & 72.1 & 4.6 & 6.8\% \\
 & 5$s$ & 19.7 & -3.6 & -15.5\% & 24.5 & 1.2 & 5.1\% & 23.3 & 24.7 & 1.4 & 6.1\% & 27.5 & 4.2 & 18.1\% \\
 & 5$p_{1/2}$ & 8.6 & -4.8 & -35.5\% & 12.6 & -0.8 & -6.3\% & 13.4 & 12.8 & -0.6 & -4.4\% & 12.9 & -0.5 & -3.6\% \\
 & 5$p_{3/2}$ & 8.0 & -4.1 & -33.9\% & 11.9 & -0.2 & -1.6\% & 12.13 & 12.1 & 0.0 & -0.3\% & 12.2 & 0.1 & 0.5\% \\
\end{tabular}
\caption{DK3 electron binding energies in eV with their absolute and relative error wrt the experiment (Ne 2$p$, Ar 3$p$, Kr 1$s$ and 4$p$, Xe 1$s$ and 4$p_{1/2}$ from Ref.~\cite{Lotz70}, the rest from Ref.~\cite{Siegbahn}).
}
\label{DK3}
\end{table*}

\bibliography{nobles_nourl}

\begin{thebibliography}{47}%
\makeatletter
\providecommand \@ifxundefined [1]{%
 \@ifx{#1\undefined}
}%
\providecommand \@ifnum [1]{%
 \ifnum #1\expandafter \@firstoftwo
 \else \expandafter \@secondoftwo
 \fi
}%
\providecommand \@ifx [1]{%
 \ifx #1\expandafter \@firstoftwo
 \else \expandafter \@secondoftwo
 \fi
}%
\providecommand \natexlab [1]{#1}%
\providecommand \enquote  [1]{``#1''}%
\providecommand \bibnamefont  [1]{#1}%
\providecommand \bibfnamefont [1]{#1}%
\providecommand \citenamefont [1]{#1}%
\providecommand \href@noop [0]{\@secondoftwo}%
\providecommand \href [0]{\begingroup \@sanitize@url \@href}%
\providecommand \@href[1]{\@@startlink{#1}\@@href}%
\providecommand \@@href[1]{\endgroup#1\@@endlink}%
\providecommand \@sanitize@url [0]{\catcode `\\12\catcode `\$12\catcode
  `\&12\catcode `\#12\catcode `\^12\catcode `\_12\catcode `\%12\relax}%
\providecommand \@@startlink[1]{}%
\providecommand \@@endlink[0]{}%
\providecommand \url  [0]{\begingroup\@sanitize@url \@url }%
\providecommand \@url [1]{\endgroup\@href {#1}{\urlprefix }}%
\providecommand \urlprefix  [0]{URL }%
\providecommand \Eprint [0]{\href }%
\providecommand \doibase [0]{http://dx.doi.org/}%
\providecommand \selectlanguage [0]{\@gobble}%
\providecommand \bibinfo  [0]{\@secondoftwo}%
\providecommand \bibfield  [0]{\@secondoftwo}%
\providecommand \translation [1]{[#1]}%
\providecommand \BibitemOpen [0]{}%
\providecommand \bibitemStop [0]{}%
\providecommand \bibitemNoStop [0]{.\EOS\space}%
\providecommand \EOS [0]{\spacefactor3000\relax}%
\providecommand \BibitemShut  [1]{\csname bibitem#1\endcsname}%
\let\auto@bib@innerbib\@empty
\bibitem [{\citenamefont {Martin}\ \emph {et~al.}(2016)\citenamefont {Martin},
  \citenamefont {Reining},\ and\ \citenamefont
  {Ceperley}}]{MartinReiningCeperley}%
  \BibitemOpen
  \bibfield  {author} {\bibinfo {author} {\bibfnamefont {R.}~\bibnamefont
  {Martin}}, \bibinfo {author} {\bibfnamefont {L.}~\bibnamefont {Reining}}, \
  and\ \bibinfo {author} {\bibfnamefont {D.~M.}\ \bibnamefont {Ceperley}},\
  }\href@noop {} {\emph {\bibinfo {title} {Interacting Electrons}}}\ (\bibinfo
  {publisher} {Cambridge University Press},\ \bibinfo {address} {Cambridge},\
  \bibinfo {year} {2016})\BibitemShut {NoStop}%
\bibitem [{\citenamefont {Siegbhan}\ \emph {et~al.}(1969)\citenamefont
  {Siegbhan} \emph {et~al.}}]{Siegbahn}%
  \BibitemOpen
  \bibfield  {author} {\bibinfo {author} {\bibfnamefont {K.}~\bibnamefont
  {Siegbhan}} \emph {et~al.},\ }\href@noop {} {\emph {\bibinfo {title} {ESCA
  applied to free molecules}}}\ (\bibinfo  {publisher} {North-Holland},\
  \bibinfo {address} {Amsterdam},\ \bibinfo {year} {1969})\BibitemShut
  {NoStop}%
\bibitem [{\citenamefont {Carlson}(1975)}]{Carlson}%
  \BibitemOpen
  \bibfield  {author} {\bibinfo {author} {\bibfnamefont {T.~A.}\ \bibnamefont
  {Carlson}},\ }\href@noop {} {\emph {\bibinfo {title} {Photoelectron and Auger
  Spectroscopy}}}\ (\bibinfo  {publisher} {Plenum},\ \bibinfo {address} {New
  York},\ \bibinfo {year} {1975})\BibitemShut {NoStop}%
\bibitem [{\citenamefont {Fetter}\ and\ \citenamefont
  {Walecka}(1971)}]{FetterWalecka}%
  \BibitemOpen
  \bibfield  {author} {\bibinfo {author} {\bibfnamefont {A.~L.}\ \bibnamefont
  {Fetter}}\ and\ \bibinfo {author} {\bibfnamefont {J.~D.}\ \bibnamefont
  {Walecka}},\ }\href@noop {} {\emph {\bibinfo {title} {Quantum Theory of
  Many-Particle Systems}}}\ (\bibinfo  {publisher} {McGraw-Hill},\ \bibinfo
  {address} {New York},\ \bibinfo {year} {1971})\BibitemShut {NoStop}%
\bibitem [{\citenamefont {Li}\ \emph {et~al.}(2017)\citenamefont {Li},
  \citenamefont {Holzmann}, \citenamefont {Duchemin}, \citenamefont {Blase},\
  and\ \citenamefont {Olevano}}]{LiOlevano17}%
  \BibitemOpen
  \bibfield  {author} {\bibinfo {author} {\bibfnamefont {J.}~\bibnamefont
  {Li}}, \bibinfo {author} {\bibfnamefont {M.}~\bibnamefont {Holzmann}},
  \bibinfo {author} {\bibfnamefont {I.}~\bibnamefont {Duchemin}}, \bibinfo
  {author} {\bibfnamefont {X.}~\bibnamefont {Blase}}, \ and\ \bibinfo {author}
  {\bibfnamefont {V.}~\bibnamefont {Olevano}},\ }\href@noop {} {\bibfield
  {journal} {\bibinfo  {journal} {Phys. Rev. Lett.}\ }\textbf {\bibinfo
  {volume} {118}},\ \bibinfo {pages} {163001} (\bibinfo {year}
  {2017})}\BibitemShut {NoStop}%
\bibitem [{\citenamefont {Li}\ \emph {et~al.}(2019{\natexlab{a}})\citenamefont
  {Li}, \citenamefont {Drummond}, \citenamefont {Schuck},\ and\ \citenamefont
  {Olevano}}]{LiOlevano19}%
  \BibitemOpen
  \bibfield  {author} {\bibinfo {author} {\bibfnamefont {J.}~\bibnamefont
  {Li}}, \bibinfo {author} {\bibfnamefont {N.~D.}\ \bibnamefont {Drummond}},
  \bibinfo {author} {\bibfnamefont {P.}~\bibnamefont {Schuck}}, \ and\ \bibinfo
  {author} {\bibfnamefont {V.}~\bibnamefont {Olevano}},\ }\href@noop {}
  {\bibfield  {journal} {\bibinfo  {journal} {SciPost Phys.}\ }\textbf
  {\bibinfo {volume} {6}},\ \bibinfo {pages} {040} (\bibinfo {year}
  {2019}{\natexlab{a}})}\BibitemShut {NoStop}%
\bibitem [{\citenamefont {Hohenberg}\ and\ \citenamefont
  {Kohn}(1964)}]{HohenbergKohn64}%
  \BibitemOpen
  \bibfield  {author} {\bibinfo {author} {\bibfnamefont {P.}~\bibnamefont
  {Hohenberg}}\ and\ \bibinfo {author} {\bibfnamefont {W.}~\bibnamefont
  {Kohn}},\ }\href@noop {} {\bibfield  {journal} {\bibinfo  {journal} {Phys.
  Rev.}\ }\textbf {\bibinfo {volume} {136}},\ \bibinfo {pages} {B864} (\bibinfo
  {year} {1964})}\BibitemShut {NoStop}%
\bibitem [{\citenamefont {Kohn}\ and\ \citenamefont {Sham}(1965)}]{KohnSham65}%
  \BibitemOpen
  \bibfield  {author} {\bibinfo {author} {\bibfnamefont {W.}~\bibnamefont
  {Kohn}}\ and\ \bibinfo {author} {\bibfnamefont {L.~J.}\ \bibnamefont
  {Sham}},\ }\href@noop {} {\bibfield  {journal} {\bibinfo  {journal} {Phys.
  Rev.}\ }\textbf {\bibinfo {volume} {140}},\ \bibinfo {pages} {A1133}
  (\bibinfo {year} {1965})}\BibitemShut {NoStop}%
\bibitem [{\citenamefont {Jones}\ and\ \citenamefont
  {Gunnarsson}(1989)}]{JonesGunnarsson89}%
  \BibitemOpen
  \bibfield  {author} {\bibinfo {author} {\bibfnamefont {R.~O.}\ \bibnamefont
  {Jones}}\ and\ \bibinfo {author} {\bibfnamefont {O.}~\bibnamefont
  {Gunnarsson}},\ }\href@noop {} {\bibfield  {journal} {\bibinfo  {journal}
  {Rev. Mod. Phys.}\ }\textbf {\bibinfo {volume} {61}},\ \bibinfo {pages} {689}
  (\bibinfo {year} {1989})}\BibitemShut {NoStop}%
\bibitem [{\citenamefont {Martin}(2004)}]{Martin}%
  \BibitemOpen
  \bibfield  {author} {\bibinfo {author} {\bibfnamefont {R.}~\bibnamefont
  {Martin}},\ }\href@noop {} {\emph {\bibinfo {title} {Electronic Structure}}}\
  (\bibinfo  {publisher} {Cambridge University Press},\ \bibinfo {address}
  {Cambridge},\ \bibinfo {year} {2004})\BibitemShut {NoStop}%
\bibitem [{\citenamefont {Gunnarsson}\ and\ \citenamefont
  {Lundqvist}(1976)}]{GunnarssonLundqvist76}%
  \BibitemOpen
  \bibfield  {author} {\bibinfo {author} {\bibfnamefont {O.}~\bibnamefont
  {Gunnarsson}}\ and\ \bibinfo {author} {\bibfnamefont {B.~I.}\ \bibnamefont
  {Lundqvist}},\ }\href@noop {} {\bibfield  {journal} {\bibinfo  {journal}
  {Phys. Rev. B}\ }\textbf {\bibinfo {volume} {13}},\ \bibinfo {pages} {4274}
  (\bibinfo {year} {1976})}\BibitemShut {NoStop}%
\bibitem [{\citenamefont {Langreth}\ and\ \citenamefont
  {Mehl}(1983)}]{LangrethMehl83}%
  \BibitemOpen
  \bibfield  {author} {\bibinfo {author} {\bibfnamefont {D.~C.}\ \bibnamefont
  {Langreth}}\ and\ \bibinfo {author} {\bibfnamefont {M.~J.}\ \bibnamefont
  {Mehl}},\ }\href@noop {} {\bibfield  {journal} {\bibinfo  {journal} {Phys.
  Rev. B}\ }\textbf {\bibinfo {volume} {28}},\ \bibinfo {pages} {1809}
  (\bibinfo {year} {1983})}\BibitemShut {NoStop}%
\bibitem [{\citenamefont {Perdew}\ \emph {et~al.}(1996)\citenamefont {Perdew},
  \citenamefont {Burke},\ and\ \citenamefont {Ernzerhof}}]{PerdewErnzerhof96}%
  \BibitemOpen
  \bibfield  {author} {\bibinfo {author} {\bibfnamefont {J.~P.}\ \bibnamefont
  {Perdew}}, \bibinfo {author} {\bibfnamefont {K.}~\bibnamefont {Burke}}, \
  and\ \bibinfo {author} {\bibfnamefont {M.}~\bibnamefont {Ernzerhof}},\
  }\href@noop {} {\bibfield  {journal} {\bibinfo  {journal} {Phys. Rev. Lett.}\
  }\textbf {\bibinfo {volume} {77}},\ \bibinfo {pages} {3865} (\bibinfo {year}
  {1996})}\BibitemShut {NoStop}%
\bibitem [{\citenamefont {Pueyo~Bellafont}\ \emph {et~al.}(2016)\citenamefont
  {Pueyo~Bellafont}, \citenamefont {Álvarez Saiz}, \citenamefont {Viñes},\
  and\ \citenamefont {Illas}}]{PueyoBellafontIllas16}%
  \BibitemOpen
  \bibfield  {author} {\bibinfo {author} {\bibfnamefont {N.}~\bibnamefont
  {Pueyo~Bellafont}}, \bibinfo {author} {\bibfnamefont {G.}~\bibnamefont
  {Álvarez Saiz}}, \bibinfo {author} {\bibfnamefont {F.}~\bibnamefont
  {Viñes}}, \ and\ \bibinfo {author} {\bibfnamefont {F.}~\bibnamefont
  {Illas}},\ }\href@noop {} {\bibfield  {journal} {\bibinfo  {journal} {Theor.
  Chem. Acc.}\ }\textbf {\bibinfo {volume} {135}},\ \bibinfo {pages} {35}
  (\bibinfo {year} {2016})}\BibitemShut {NoStop}%
\bibitem [{\citenamefont {Golze}\ \emph {et~al.}(2018)\citenamefont {Golze},
  \citenamefont {Wilhelm}, \citenamefont {van Setten},\ and\ \citenamefont
  {Rinke}}]{GolzeRinke18}%
  \BibitemOpen
  \bibfield  {author} {\bibinfo {author} {\bibfnamefont {D.}~\bibnamefont
  {Golze}}, \bibinfo {author} {\bibfnamefont {J.}~\bibnamefont {Wilhelm}},
  \bibinfo {author} {\bibfnamefont {M.~J.}\ \bibnamefont {van Setten}}, \ and\
  \bibinfo {author} {\bibfnamefont {P.}~\bibnamefont {Rinke}},\ }\href@noop {}
  {\bibfield  {journal} {\bibinfo  {journal} {J. Chem. Theory Comput.}\
  }\textbf {\bibinfo {volume} {14}},\ \bibinfo {pages} {4856} (\bibinfo {year}
  {2018})}\BibitemShut {NoStop}%
\bibitem [{\citenamefont {Perdew}\ \emph {et~al.}(1982)\citenamefont {Perdew},
  \citenamefont {Parr}, \citenamefont {Levy},\ and\ \citenamefont
  {Balduz}}]{PerdewBalduz82}%
  \BibitemOpen
  \bibfield  {author} {\bibinfo {author} {\bibfnamefont {J.~P.}\ \bibnamefont
  {Perdew}}, \bibinfo {author} {\bibfnamefont {R.~G.}\ \bibnamefont {Parr}},
  \bibinfo {author} {\bibfnamefont {M.}~\bibnamefont {Levy}}, \ and\ \bibinfo
  {author} {\bibfnamefont {J.~L.}\ \bibnamefont {Balduz}},\ }\href@noop {}
  {\bibfield  {journal} {\bibinfo  {journal} {Phys. Rev. Lett.}\ }\textbf
  {\bibinfo {volume} {49}},\ \bibinfo {pages} {1691} (\bibinfo {year}
  {1982})}\BibitemShut {NoStop}%
\bibitem [{\citenamefont {Levy}\ \emph {et~al.}(1984)\citenamefont {Levy},
  \citenamefont {Perdew},\ and\ \citenamefont {Sahni}}]{LevySahni84}%
  \BibitemOpen
  \bibfield  {author} {\bibinfo {author} {\bibfnamefont {M.}~\bibnamefont
  {Levy}}, \bibinfo {author} {\bibfnamefont {J.~P.}\ \bibnamefont {Perdew}}, \
  and\ \bibinfo {author} {\bibfnamefont {V.}~\bibnamefont {Sahni}},\
  }\href@noop {} {\bibfield  {journal} {\bibinfo  {journal} {Phys. Rev. A}\
  }\textbf {\bibinfo {volume} {30}},\ \bibinfo {pages} {2745} (\bibinfo {year}
  {1984})}\BibitemShut {NoStop}%
\bibitem [{\citenamefont {Almbladh}\ and\ \citenamefont {von
  Barth}(1985)}]{AlmbladhVonBarth85}%
  \BibitemOpen
  \bibfield  {author} {\bibinfo {author} {\bibfnamefont {C.-O.}\ \bibnamefont
  {Almbladh}}\ and\ \bibinfo {author} {\bibfnamefont {U.}~\bibnamefont {von
  Barth}},\ }\href@noop {} {\bibfield  {journal} {\bibinfo  {journal} {Phys.
  Rev. B}\ }\textbf {\bibinfo {volume} {31}},\ \bibinfo {pages} {3231}
  (\bibinfo {year} {1985})}\BibitemShut {NoStop}%
\bibitem [{\citenamefont {Hedin}(1965)}]{Hedin65}%
  \BibitemOpen
  \bibfield  {author} {\bibinfo {author} {\bibfnamefont {L.}~\bibnamefont
  {Hedin}},\ }\href@noop {} {\bibfield  {journal} {\bibinfo  {journal} {Phys.
  Rev.}\ }\textbf {\bibinfo {volume} {139}},\ \bibinfo {pages} {A796} (\bibinfo
  {year} {1965})}\BibitemShut {NoStop}%
\bibitem [{\citenamefont {Strinati}\ \emph {et~al.}(1980)\citenamefont
  {Strinati}, \citenamefont {Mattausch},\ and\ \citenamefont
  {Hanke}}]{StrinatiHanke80}%
  \BibitemOpen
  \bibfield  {author} {\bibinfo {author} {\bibfnamefont {G.}~\bibnamefont
  {Strinati}}, \bibinfo {author} {\bibfnamefont {H.~J.}\ \bibnamefont
  {Mattausch}}, \ and\ \bibinfo {author} {\bibfnamefont {W.}~\bibnamefont
  {Hanke}},\ }\href@noop {} {\bibfield  {journal} {\bibinfo  {journal} {Phys.
  Rev. Lett.}\ }\textbf {\bibinfo {volume} {45}},\ \bibinfo {pages} {290}
  (\bibinfo {year} {1980})}\BibitemShut {NoStop}%
\bibitem [{\citenamefont {Strinati}\ \emph {et~al.}(1982)\citenamefont
  {Strinati}, \citenamefont {Mattausch},\ and\ \citenamefont
  {Hanke}}]{StrinatiHanke82}%
  \BibitemOpen
  \bibfield  {author} {\bibinfo {author} {\bibfnamefont {G.}~\bibnamefont
  {Strinati}}, \bibinfo {author} {\bibfnamefont {H.~J.}\ \bibnamefont
  {Mattausch}}, \ and\ \bibinfo {author} {\bibfnamefont {W.}~\bibnamefont
  {Hanke}},\ }\href@noop {} {\bibfield  {journal} {\bibinfo  {journal} {Phys.
  Rev. B}\ }\textbf {\bibinfo {volume} {25}},\ \bibinfo {pages} {2867}
  (\bibinfo {year} {1982})}\BibitemShut {NoStop}%
\bibitem [{\citenamefont {Hybertsen}\ and\ \citenamefont
  {Louie}(1985)}]{HybertsenLouie85}%
  \BibitemOpen
  \bibfield  {author} {\bibinfo {author} {\bibfnamefont {M.~S.}\ \bibnamefont
  {Hybertsen}}\ and\ \bibinfo {author} {\bibfnamefont {S.~G.}\ \bibnamefont
  {Louie}},\ }\href@noop {} {\bibfield  {journal} {\bibinfo  {journal} {Phys.
  Rev. Lett.}\ }\textbf {\bibinfo {volume} {55}},\ \bibinfo {pages} {1418}
  (\bibinfo {year} {1985})}\BibitemShut {NoStop}%
\bibitem [{\citenamefont {Godby}\ \emph {et~al.}(1987)\citenamefont {Godby},
  \citenamefont {Schl{\"u}ter},\ and\ \citenamefont {Sham}}]{GodbySham87}%
  \BibitemOpen
  \bibfield  {author} {\bibinfo {author} {\bibfnamefont {R.~W.}\ \bibnamefont
  {Godby}}, \bibinfo {author} {\bibfnamefont {M.}~\bibnamefont {Schl{\"u}ter}},
  \ and\ \bibinfo {author} {\bibfnamefont {L.~J.}\ \bibnamefont {Sham}},\
  }\href@noop {} {\bibfield  {journal} {\bibinfo  {journal} {Phys. Rev. B}\
  }\textbf {\bibinfo {volume} {35}},\ \bibinfo {pages} {4170} (\bibinfo {year}
  {1987})}\BibitemShut {NoStop}%
\bibitem [{\citenamefont {Hedin}(1999)}]{Hedin99}%
  \BibitemOpen
  \bibfield  {author} {\bibinfo {author} {\bibfnamefont {L.}~\bibnamefont
  {Hedin}},\ }\href@noop {} {\bibfield  {journal} {\bibinfo  {journal} {J.
  Phys. Condens. Matter}\ }\textbf {\bibinfo {volume} {11}},\ \bibinfo {pages}
  {R489} (\bibinfo {year} {1999})}\BibitemShut {NoStop}%
\bibitem [{\citenamefont {Li}\ \emph {et~al.}(2019{\natexlab{b}})\citenamefont
  {Li}, \citenamefont {Duchemin}, \citenamefont {Roscioni}, \citenamefont
  {Friederich}, \citenamefont {Anderson}, \citenamefont {Da~Como},
  \citenamefont {Kociok-Köhn}, \citenamefont {Wenzel}, \citenamefont
  {Zannoni}, \citenamefont {Beljonne}, \citenamefont {Blase},\ and\
  \citenamefont {D’Avino}}]{Li2019}%
  \BibitemOpen
  \bibfield  {author} {\bibinfo {author} {\bibfnamefont {J.}~\bibnamefont
  {Li}}, \bibinfo {author} {\bibfnamefont {I.}~\bibnamefont {Duchemin}},
  \bibinfo {author} {\bibfnamefont {O.~M.}\ \bibnamefont {Roscioni}}, \bibinfo
  {author} {\bibfnamefont {P.}~\bibnamefont {Friederich}}, \bibinfo {author}
  {\bibfnamefont {M.}~\bibnamefont {Anderson}}, \bibinfo {author}
  {\bibfnamefont {E.}~\bibnamefont {Da~Como}}, \bibinfo {author} {\bibfnamefont
  {G.}~\bibnamefont {Kociok-Köhn}}, \bibinfo {author} {\bibfnamefont
  {W.}~\bibnamefont {Wenzel}}, \bibinfo {author} {\bibfnamefont
  {C.}~\bibnamefont {Zannoni}}, \bibinfo {author} {\bibfnamefont
  {D.}~\bibnamefont {Beljonne}}, \bibinfo {author} {\bibfnamefont
  {X.}~\bibnamefont {Blase}}, \ and\ \bibinfo {author} {\bibfnamefont
  {G.}~\bibnamefont {D’Avino}},\ }\href@noop {} {\bibfield  {journal}
  {\bibinfo  {journal} {Materials Horizons}\ }\textbf {\bibinfo {volume} {6}},\
  \bibinfo {pages} {107} (\bibinfo {year} {2019}{\natexlab{b}})}\BibitemShut
  {NoStop}%
\bibitem [{\citenamefont {Bruneval}(2012)}]{Bruneval12}%
  \BibitemOpen
  \bibfield  {author} {\bibinfo {author} {\bibfnamefont {F.}~\bibnamefont
  {Bruneval}},\ }\href@noop {} {\bibfield  {journal} {\bibinfo  {journal} {J.
  Chem. Phys.}\ }\textbf {\bibinfo {volume} {136}},\ \bibinfo {pages} {194107}
  (\bibinfo {year} {2012})}\BibitemShut {NoStop}%
\bibitem [{\citenamefont {van Setten}\ \emph {et~al.}(2015)\citenamefont {van
  Setten} \emph {et~al.}}]{VanSettenRinke15}%
  \BibitemOpen
  \bibfield  {author} {\bibinfo {author} {\bibfnamefont {M.~J.}\ \bibnamefont
  {van Setten}} \emph {et~al.},\ }\href@noop {} {\bibfield  {journal} {\bibinfo
   {journal} {J. Chem. Theory Comput.}\ }\textbf {\bibinfo {volume} {11}},\
  \bibinfo {pages} {5665} (\bibinfo {year} {2015})}\BibitemShut {NoStop}%
\bibitem [{\citenamefont {Lotz}(1970)}]{Lotz70}%
  \BibitemOpen
  \bibfield  {author} {\bibinfo {author} {\bibfnamefont {W.}~\bibnamefont
  {Lotz}},\ }\href@noop {} {\bibfield  {journal} {\bibinfo  {journal} {J. Opt.
  Soc. Am.}\ }\textbf {\bibinfo {volume} {60}},\ \bibinfo {pages} {206}
  (\bibinfo {year} {1970})}\BibitemShut {NoStop}%
\bibitem [{\citenamefont {Chang}\ \emph {et~al.}(1986)\citenamefont {Chang},
  \citenamefont {P\'elissier},\ and\ \citenamefont {Durand}}]{ChangDurand86}%
  \BibitemOpen
  \bibfield  {author} {\bibinfo {author} {\bibfnamefont {C.}~\bibnamefont
  {Chang}}, \bibinfo {author} {\bibfnamefont {M.}~\bibnamefont {P\'elissier}},
  \ and\ \bibinfo {author} {\bibfnamefont {P.}~\bibnamefont {Durand}},\
  }\href@noop {} {\bibfield  {journal} {\bibinfo  {journal} {Phys. Scr.}\
  }\textbf {\bibinfo {volume} {34}},\ \bibinfo {pages} {394} (\bibinfo {year}
  {1986})}\BibitemShut {NoStop}%
\bibitem [{\citenamefont {Heully}\ \emph {et~al.}(1986)\citenamefont {Heully},
  \citenamefont {Lindgren}, \citenamefont {Lindroth}, \citenamefont
  {Lundqvist},\ and\ \citenamefont
  {Martensson-Pendrill}}]{HeullyMartensson-Pendrill86}%
  \BibitemOpen
  \bibfield  {author} {\bibinfo {author} {\bibfnamefont {J.-L.}\ \bibnamefont
  {Heully}}, \bibinfo {author} {\bibfnamefont {I.}~\bibnamefont {Lindgren}},
  \bibinfo {author} {\bibfnamefont {E.}~\bibnamefont {Lindroth}}, \bibinfo
  {author} {\bibfnamefont {S.}~\bibnamefont {Lundqvist}}, \ and\ \bibinfo
  {author} {\bibfnamefont {A.-M.}\ \bibnamefont {Martensson-Pendrill}},\
  }\href@noop {} {\bibfield  {journal} {\bibinfo  {journal} {J. Phys. B}\
  }\textbf {\bibinfo {volume} {19}},\ \bibinfo {pages} {2799} (\bibinfo {year}
  {1986})}\BibitemShut {NoStop}%
\bibitem [{\citenamefont {van Lenthe}\ \emph {et~al.}(1993)\citenamefont {van
  Lenthe}, \citenamefont {Baerends},\ and\ \citenamefont
  {Snijders}}]{vanLentheSnijders93}%
  \BibitemOpen
  \bibfield  {author} {\bibinfo {author} {\bibfnamefont {E.}~\bibnamefont {van
  Lenthe}}, \bibinfo {author} {\bibfnamefont {E.~J.}\ \bibnamefont {Baerends}},
  \ and\ \bibinfo {author} {\bibfnamefont {J.~G.}\ \bibnamefont {Snijders}},\
  }\href@noop {} {\bibfield  {journal} {\bibinfo  {journal} {J. Chem. Phys.}\
  }\textbf {\bibinfo {volume} {99}},\ \bibinfo {pages} {4597} (\bibinfo {year}
  {1993})}\BibitemShut {NoStop}%
\bibitem [{\citenamefont {Nakajima}\ and\ \citenamefont
  {Hirao}(2000{\natexlab{a}})}]{NakajimaHirao00}%
  \BibitemOpen
  \bibfield  {author} {\bibinfo {author} {\bibfnamefont {T.}~\bibnamefont
  {Nakajima}}\ and\ \bibinfo {author} {\bibfnamefont {K.}~\bibnamefont
  {Hirao}},\ }\href@noop {} {\bibfield  {journal} {\bibinfo  {journal} {Chem.
  Phys. Lett.}\ }\textbf {\bibinfo {volume} {329}},\ \bibinfo {pages} {511}
  (\bibinfo {year} {2000}{\natexlab{a}})}\BibitemShut {NoStop}%
\bibitem [{\citenamefont {Nakajima}\ and\ \citenamefont
  {Hirao}(2000{\natexlab{b}})}]{NakajimaHirao00b}%
  \BibitemOpen
  \bibfield  {author} {\bibinfo {author} {\bibfnamefont {T.}~\bibnamefont
  {Nakajima}}\ and\ \bibinfo {author} {\bibfnamefont {K.}~\bibnamefont
  {Hirao}},\ }\href@noop {} {\bibfield  {journal} {\bibinfo  {journal} {J.
  Chem. Phys.}\ }\textbf {\bibinfo {volume} {113}},\ \bibinfo {pages} {7786}
  (\bibinfo {year} {2000}{\natexlab{b}})}\BibitemShut {NoStop}%
\bibitem [{\citenamefont {Pollak}\ and\ \citenamefont
  {Weigend}(2017)}]{Pollak_2017}%
  \BibitemOpen
  \bibfield  {author} {\bibinfo {author} {\bibfnamefont {P.}~\bibnamefont
  {Pollak}}\ and\ \bibinfo {author} {\bibfnamefont {F.}~\bibnamefont
  {Weigend}},\ }\href@noop {} {\bibfield  {journal} {\bibinfo  {journal} {J.
  Chem. Theory Comput.}\ }\textbf {\bibinfo {volume} {13}},\ \bibinfo {pages}
  {3696} (\bibinfo {year} {2017})}\BibitemShut {NoStop}%
\bibitem [{\citenamefont {Duchemin}\ \emph {et~al.}(2017)\citenamefont
  {Duchemin}, \citenamefont {Li},\ and\ \citenamefont {Blase}}]{Duchemin_2017}%
  \BibitemOpen
  \bibfield  {author} {\bibinfo {author} {\bibfnamefont {I.}~\bibnamefont
  {Duchemin}}, \bibinfo {author} {\bibfnamefont {J.}~\bibnamefont {Li}}, \ and\
  \bibinfo {author} {\bibfnamefont {X.}~\bibnamefont {Blase}},\ }\href@noop {}
  {\bibfield  {journal} {\bibinfo  {journal} {J. Chem. Theory Comput.}\
  }\textbf {\bibinfo {volume} {13}},\ \bibinfo {pages} {1199} (\bibinfo {year}
  {2017})}\BibitemShut {NoStop}%
\bibitem [{\citenamefont {Weigend}(2008)}]{Weigend_2008}%
  \BibitemOpen
  \bibfield  {author} {\bibinfo {author} {\bibfnamefont {F.}~\bibnamefont
  {Weigend}},\ }\href@noop {} {\bibfield  {journal} {\bibinfo  {journal} {J.
  Comput. Chem.}\ }\textbf {\bibinfo {volume} {29}},\ \bibinfo {pages} {167}
  (\bibinfo {year} {2008})}\BibitemShut {NoStop}%
\bibitem [{\citenamefont {Stoychev}\ \emph {et~al.}(2017)\citenamefont
  {Stoychev}, \citenamefont {Auer},\ and\ \citenamefont
  {Neese}}]{Stoychev_2017}%
  \BibitemOpen
  \bibfield  {author} {\bibinfo {author} {\bibfnamefont {G.~L.}\ \bibnamefont
  {Stoychev}}, \bibinfo {author} {\bibfnamefont {A.~A.}\ \bibnamefont {Auer}},
  \ and\ \bibinfo {author} {\bibfnamefont {F.}~\bibnamefont {Neese}},\
  }\href@noop {} {\bibfield  {journal} {\bibinfo  {journal} {J. Chem. Theory
  Comput.}\ }\textbf {\bibinfo {volume} {13}},\ \bibinfo {pages} {554}
  (\bibinfo {year} {2017})}\BibitemShut {NoStop}%
\bibitem [{\citenamefont {Valiev}\ \emph {et~al.}(2010)\citenamefont {Valiev},
  \citenamefont {Bylaska}, \citenamefont {Govind}, \citenamefont {Kowalski},
  \citenamefont {Straatsma}, \citenamefont {Dam}, \citenamefont {Wang},
  \citenamefont {Nieplocha}, \citenamefont {Apra}, \citenamefont {Windus},\
  and\ \citenamefont {de~Jong}}]{nwchem}%
  \BibitemOpen
  \bibfield  {author} {\bibinfo {author} {\bibfnamefont {M.}~\bibnamefont
  {Valiev}}, \bibinfo {author} {\bibfnamefont {E.}~\bibnamefont {Bylaska}},
  \bibinfo {author} {\bibfnamefont {N.}~\bibnamefont {Govind}}, \bibinfo
  {author} {\bibfnamefont {K.}~\bibnamefont {Kowalski}}, \bibinfo {author}
  {\bibfnamefont {T.}~\bibnamefont {Straatsma}}, \bibinfo {author}
  {\bibfnamefont {H.~V.}\ \bibnamefont {Dam}}, \bibinfo {author} {\bibfnamefont
  {D.}~\bibnamefont {Wang}}, \bibinfo {author} {\bibfnamefont {J.}~\bibnamefont
  {Nieplocha}}, \bibinfo {author} {\bibfnamefont {E.}~\bibnamefont {Apra}},
  \bibinfo {author} {\bibfnamefont {T.}~\bibnamefont {Windus}}, \ and\ \bibinfo
  {author} {\bibfnamefont {W.}~\bibnamefont {de~Jong}},\ }\href@noop {}
  {\bibfield  {journal} {\bibinfo  {journal} {Comput. Phys. Commun.}\ }\textbf
  {\bibinfo {volume} {181}},\ \bibinfo {pages} {1477 } (\bibinfo {year}
  {2010})}\BibitemShut {NoStop}%
\bibitem [{\citenamefont {Blase}\ \emph {et~al.}(2011)\citenamefont {Blase},
  \citenamefont {Attaccalite},\ and\ \citenamefont {Olevano}}]{BlaseOlevano11}%
  \BibitemOpen
  \bibfield  {author} {\bibinfo {author} {\bibfnamefont {X.}~\bibnamefont
  {Blase}}, \bibinfo {author} {\bibfnamefont {C.}~\bibnamefont {Attaccalite}},
  \ and\ \bibinfo {author} {\bibfnamefont {V.}~\bibnamefont {Olevano}},\
  }\href@noop {} {\bibfield  {journal} {\bibinfo  {journal} {Phys. Rev. B}\
  }\textbf {\bibinfo {volume} {83}},\ \bibinfo {pages} {115103} (\bibinfo
  {year} {2011})}\BibitemShut {NoStop}%
\bibitem [{\citenamefont {Jacquemin}\ \emph {et~al.}(2015)\citenamefont
  {Jacquemin}, \citenamefont {Duchemin},\ and\ \citenamefont {Blase}}]{Jac15a}%
  \BibitemOpen
  \bibfield  {author} {\bibinfo {author} {\bibfnamefont {D.}~\bibnamefont
  {Jacquemin}}, \bibinfo {author} {\bibfnamefont {I.}~\bibnamefont {Duchemin}},
  \ and\ \bibinfo {author} {\bibfnamefont {X.}~\bibnamefont {Blase}},\
  }\href@noop {} {\bibfield  {journal} {\bibinfo  {journal} {J. Chem. Theory
  Comput.}\ }\textbf {\bibinfo {volume} {11}},\ \bibinfo {pages} {3290}
  (\bibinfo {year} {2015})}\BibitemShut {NoStop}%
\bibitem [{\citenamefont {Li}\ \emph {et~al.}(2016)\citenamefont {Li},
  \citenamefont {D'Avino}, \citenamefont {Duchemin}, \citenamefont {Beljonne},\
  and\ \citenamefont {Blase}}]{Li16}%
  \BibitemOpen
  \bibfield  {author} {\bibinfo {author} {\bibfnamefont {J.}~\bibnamefont
  {Li}}, \bibinfo {author} {\bibfnamefont {G.}~\bibnamefont {D'Avino}},
  \bibinfo {author} {\bibfnamefont {I.}~\bibnamefont {Duchemin}}, \bibinfo
  {author} {\bibfnamefont {D.}~\bibnamefont {Beljonne}}, \ and\ \bibinfo
  {author} {\bibfnamefont {X.}~\bibnamefont {Blase}},\ }\href@noop {}
  {\bibfield  {journal} {\bibinfo  {journal} {J. Phys. Chem. Lett.}\ }\textbf
  {\bibinfo {volume} {7}},\ \bibinfo {pages} {2814} (\bibinfo {year}
  {2016})}\BibitemShut {NoStop}%
\bibitem [{\citenamefont {Balasubramani}\ \emph {et~al.}(2020)\citenamefont
  {Balasubramani} \emph {et~al.}}]{turbomole}%
  \BibitemOpen
  \bibfield  {author} {\bibinfo {author} {\bibfnamefont {S.~G.}\ \bibnamefont
  {Balasubramani}} \emph {et~al.},\ }\href@noop {} {\bibfield  {journal}
  {\bibinfo  {journal} {J. Chem. Phys.}\ }\textbf {\bibinfo {volume} {152}},\
  \bibinfo {pages} {184107} (\bibinfo {year} {2020})}\BibitemShut {NoStop}%
\bibitem [{\citenamefont {Thompson}\ \emph {et~al.}(2009)\citenamefont
  {Thompson} \emph {et~al.}}]{Williams}%
  \BibitemOpen
  \bibfield  {author} {\bibinfo {author} {\bibfnamefont {A.~C.}\ \bibnamefont
  {Thompson}} \emph {et~al.},\ }\href@noop {} {\emph {\bibinfo {title} {X-ray
  Data Booklet}}}\ (\bibinfo  {publisher} {Lawrence Berkeley Nat. Lab},\
  \bibinfo {address} {Berkeley},\ \bibinfo {year} {2009})\BibitemShut {NoStop}%
\bibitem [{\citenamefont {Bearden}\ and\ \citenamefont
  {Burr}(1967)}]{BeardenBurr67}%
  \BibitemOpen
  \bibfield  {author} {\bibinfo {author} {\bibfnamefont {J.~A.}\ \bibnamefont
  {Bearden}}\ and\ \bibinfo {author} {\bibfnamefont {A.~F.}\ \bibnamefont
  {Burr}},\ }\href@noop {} {\bibfield  {journal} {\bibinfo  {journal} {Rev.
  Mod. Phys.}\ }\textbf {\bibinfo {volume} {39}},\ \bibinfo {pages} {125}
  (\bibinfo {year} {1967})}\BibitemShut {NoStop}%
\bibitem [{\citenamefont {Pia}\ \emph {et~al.}(2011)\citenamefont {Pia},
  \citenamefont {Seo}, \citenamefont {Batic}, \citenamefont {Begalli},
  \citenamefont {Kim}, \citenamefont {L.},\ and\ \citenamefont
  {Saracco}}]{PiaSaracco11}%
  \BibitemOpen
  \bibfield  {author} {\bibinfo {author} {\bibfnamefont {M.~G.}\ \bibnamefont
  {Pia}}, \bibinfo {author} {\bibfnamefont {H.}~\bibnamefont {Seo}}, \bibinfo
  {author} {\bibfnamefont {M.}~\bibnamefont {Batic}}, \bibinfo {author}
  {\bibfnamefont {M.}~\bibnamefont {Begalli}}, \bibinfo {author} {\bibfnamefont
  {C.~H.}\ \bibnamefont {Kim}}, \bibinfo {author} {\bibfnamefont
  {Q.}~\bibnamefont {L.}}, \ and\ \bibinfo {author} {\bibfnamefont
  {P.}~\bibnamefont {Saracco}},\ }\href@noop {} {\bibfield  {journal} {\bibinfo
   {journal} {IEEE Trans. Nucl. Science}\ }\textbf {\bibinfo {volume} {58}},\
  \bibinfo {pages} {3246} (\bibinfo {year} {2011})}\BibitemShut {NoStop}%
\bibitem [{\citenamefont {Atalla}\ \emph {et~al.}(2013)\citenamefont {Atalla},
  \citenamefont {Yoon}, \citenamefont {Caruso}, \citenamefont {Rinke},\ and\
  \citenamefont {Scheffler}}]{AtallaScheffler13}%
  \BibitemOpen
  \bibfield  {author} {\bibinfo {author} {\bibfnamefont {V.}~\bibnamefont
  {Atalla}}, \bibinfo {author} {\bibfnamefont {M.}~\bibnamefont {Yoon}},
  \bibinfo {author} {\bibfnamefont {F.}~\bibnamefont {Caruso}}, \bibinfo
  {author} {\bibfnamefont {P.}~\bibnamefont {Rinke}}, \ and\ \bibinfo {author}
  {\bibfnamefont {M.}~\bibnamefont {Scheffler}},\ }\href@noop {} {\bibfield
  {journal} {\bibinfo  {journal} {Phys. Rev. B}\ }\textbf {\bibinfo {volume}
  {88}},\ \bibinfo {pages} {165122} (\bibinfo {year} {2013})}\BibitemShut
  {NoStop}%
\bibitem [{\citenamefont {Olevano}\ and\ \citenamefont
  {Ladisa}(2010)}]{OlevanoLadisa10}%
  \BibitemOpen
  \bibfield  {author} {\bibinfo {author} {\bibfnamefont {V.}~\bibnamefont
  {Olevano}}\ and\ \bibinfo {author} {\bibfnamefont {M.}~\bibnamefont
  {Ladisa}},\ }\href@noop {} {\enquote {\bibinfo {title} {Condensed matter
  many-body theory in relativistic covariant form},}\ } (\bibinfo {year}
  {2010}),\ \Eprint {http://arxiv.org/abs/1002.2117} {arXiv:1002.2117}
  \BibitemShut {NoStop}%
\end{thebibliography}%

\end{document}